\documentstyle[tighten,preprint,pra,aps]{revtex}
\newcommand{\eref}[1]{(\ref{#1})}
\begin{document}

\draft
\title{Enhancement of the electric dipole moment of the electron
 in BaF molecule.}
\author{M. G. Kozlov, A. V. Titov, N. S. Mosyagin, and P. V. Souchko}
\address{Petersburg Nuclear Physics Institute, \\
         Gatchina, St.-Petersburg district 188350, RUSSIA}
\date{10 July 1997}
\maketitle

\begin{abstract}
We report results of {\it ab initio} calculation of the spin-rotational
Hamiltonian parameters including $P$- and $P,T$-odd terms for the BaF
molecule. The ground state wave function of BaF molecule is found with the
help of the Relativistic Effective Core Potential method followed by the
restoration of molecular four-component spinors in the core region of
barium in the framework of a non-variational procedure. Core polarization
effects are included with the help of the atomic Many Body Perturbation
Theory for Barium atom. For the hyperfine constants the accuracy of this
method is about 5-10\%.

\end{abstract}


%*************************************************************************

\narrowtext
\paragraph{Introduction.}

It is well known that possible $P$- and $P,T$-odd effects are strongly
enhanced in heavy diatomic radicals (see, for example, \cite{IBK,KL}).  In
the molecular experiment with the TlF molecule \cite{CSH} stringent limits
on the Schiff moment of the Tl nucleus and on the tensor constant of the
electron-nucleus $P,T$-odd interaction were obtained. In the experiments
with the polar diatomics with the unpaired electron one can search for the
$P,T$-odd effects caused by the permanent electric dipole moment (EDM) of
the electron $d_{\rm e}$ \cite{SF} and by the scalar electron-nuclear
$P,T$-odd interaction \cite{GLM}. The most stringent limit on the electron
EDM was obtained in the experiment with atomic Thallium
\cite{CRDR} (for the review of the theoretical predictions for $d_{\rm e}$
see \cite{Barr}).  Heavy polar diatomic molecules provide enhancement of
the electron EDM, which is several orders of magnitude larger, than
in Tl.  An experimental search for the EDM of the electron is now underway
on the YbF molecule \cite{SPRH}.  The $P$-odd effects associated with the
anapole moment of the nucleus are also strongly enhanced in diatomic
radicals \cite{L,FK}.

The first calculations of the $P,T$-odd interactions in molecules were
carried out for TlF molecule with the use of a ``relativistic matching''
of nonrelativistic one-configurational wave function \cite{Hinds3}.

Then a semi-empirical scheme~\cite{K85,KE} and {\it ab initio}
method based on the Relativistic Effective Core Potential (RECP)
calculation of the molecular wave function~\cite{Dmitriev} were developed.
The first RECP-based calculations of the $P,T$-odd spin-rotational
Hamiltonian parameters for PbF and HgF molecules were carried out in the
framework of the one-configurational approximation
with minimal atomic basis sets, i.e.\ the correlation structure was not
taken into account. In calculation of YbF molecule~\cite{TME}, a
flexible atomic basis set was used and the correlation effects were
considered within the Restricted Active Space SCF (RASSCF)
method~\cite{RASSCF,MOLCAS}.

It was concluded in \cite{TME} that in order to perform more accurate
calculations of the hyperfine and the $P,T$-odd constants, the
spin-correlation of the unpaired electron with the outermost core shells
$5s$ and $5p$ of ytterbium should be taken into account.  Such correlations
can be hardly efficiently considered within MC~SCF-like methods because of
the necessity to correlate too many electrons.

Here we suggest to use an effective operator (EO) technique to account for
the most important types of the core-valence correlations. EOs
for the valence electrons are formed with the help of the atomic
many body perturbation theory. This method allows to include
correlations not only with the outermost core shells, but with all core
electrons, which appears to be quite important for the hyperfine and
$P,T$-odd interactions.  The EO technique was recently developed for atoms
\cite{DFK} and proved to be very efficient for the calculations of the
hyperfine structure of the heavy atoms \cite{DFKP}.  This technique is
naturally and easily combined with the RECP method for the molecular
calculations. As a result, a significant improvement of the accuracy is
achieved.

Below we report the results of application of this method to
calculation of the BaF molecule.

\paragraph{Spin-rotational Hamiltonian.}

Molecular spin-rota\-tional degrees of freedom are described by the
following spin-rotational Hamiltonian (see \cite{KL}):
%----------------------------------------------------------------------
\begin{eqnarray}
        H_{\rm sr} & = & B{\bf N}^2 + \gamma {\bf S N} -D_e{\bf n E}
                 + {\bf S \hat{A} I}
\nonumber\\
                   & + & W_{\rm A} k_{\rm A} {\bf n \times S \cdot I}
                         +(W_{\rm S} k_{\rm S}
                         + W_d d_{\rm e}) {\bf S n}.
\label{1}
\end{eqnarray}
%----------------------------------------------------------------------
In this expression
$\bf N$ is the rotational angular momentum, $B$ is the rotational constant,
$\bf S$ and $\bf I$ are the spins of the electron and the Ba nucleus,
$\bf n$ is the unit vector directed along the molecular axis from Ba to F.
The spin-doubling constant $\gamma$ characterizes the spin-rotational
interaction. $D_e$ and $\bf E$ are the molecular dipole moment and the
external electric field.
The axial tensor $\bf \hat{A}$ describes magnetic
hyperfine structure. It can be determined by two parameters:
$A=(A_{\parallel}+2A_{\perp})/3$ and $A_{\rm
d}=(A_{\parallel}-A_{\perp})/3$.
The last three terms in \eref{1} account for the $P$- and $P,T$-odd
effects.  First of them describes interaction of the electron spin with the
anapole moment of the nucleus $k_{\rm A}$ \cite{FK}. The second one
corresponds to the scalar $P,T$-odd electron-nucleus interaction with the
dimensionless constant $k_{\rm S}$. The third one describes interaction of
the electron EDM $d_{\rm e}$ with the molecular field. Constant $W_d$
characterizes an effective electric field on the unpaired electron.

It is important to note that all $P$- and $P,T$-odd constants $W_i$ mostly
depend on the electron spin-density in the vicinity of the heavy nucleus.
The same, of course, can be said about hyperfine constants $A$ and
$A_{\rm d}$. So, the comparison of the theoretical results for the hyperfine
constants with the experiment is a good test for the accuracy of the whole
calculation.

%*************************************************************************
\paragraph{RECP calculation of electronic wave function.}
\label{S3}

The scheme of the RECP calculation for BaF molecule is very similar to that
for YbF described in~\cite{TME} (see also~\cite{Dmitriev})
%### due to close electronic structure in valence region
and below we will focus only on
specific features of the present calculations.

%The Generalized RECP (GRECP)~\cite{Tupitsyn1} with
%  $1s^2 [\ldots] 4s^2 4p^6 4d^{10}$
%shells treated as inner core (and not included explicitly in the
%RECP calculations) was se\-lec\-ted for calculations of BaF from a few other
%RECP variants generated in this work because it provided high accuracy for
%the spin-rotational Hamiltonian parameters with quite small computational
%expenses (see table~\ref{Ba_conf}).

The Generalized RECP (GRECP)~\cite{Tupitsyn1} (with the inner core $1s^2
[\ldots] 4s^2 4p^6 4d^{10}$ shells which were not included explicitly in
the RECP calculations) was selected from a few other RECP variants for
calculations of BaF because our test electronic structure calculations
showed that it combined high accuracy with quite small computational
expenses (see table~\ref{Ba_conf} and the spectroscopic data below).

Numerical pseudospinors derived from the GRECP/SCF calculations of some
electronic configurations for Ba, Ba$^+$ and Ba$^{++}$ were approximated
by generally contracted $s,p,d$ and $f$ gaussian functions forming
 $(10,8,6,2) \rightarrow [6,5,4,2]$  basis set for
%-------------
barium\footnote{See \cite{TME} for details. Gaussian expansions for these
pseudo\-spi\-nors, GRECP com\-po\-nents and MO~LCAO coefficients from BaF
calculations can be found on {\bf http://www.qchem.pnpi.spb.ru}.}.
%-------------
For fluorine we used basis sets $(14,9,4) \rightarrow [6,5,2]$ and
 $[4,3,3]$ from the ANO-I Library~\cite{MOLCAS}.  These basis sets proved
to be sufficiently flexible to reproduce electronic structure in valence
region of BaF as compared to other basis sets involved in our test SCF and
RASSCF calculations.

The RASSCF calculations of the spectroscopic constants were performed with
the spin-Averaged part of the GRECP (AREP) and contribution of relatively
small spin-orbit interaction (i.e.\ Effective Spin-Orbit Potential or ESOP
as a part of GRECP) was estimated in the framework of the perturbation
theory.
The results of our AREP/RASSCF calculations with 79558 configurations
%-------------
\footnote{We used $C_{2v}$ point group with ($a_1,b_1,b_2,a_2$) irreducible
representations; 17 electrons were distributed on active orbitals
within RAS~1=(3,1,1,0), RAS~2=(3,1,1,0) and RAS~3=(5,3,3,1) subspaces.}
%-------------
for the equilibrium distance and vibration constant
($R_e = 2.25~\AA$, $\omega_e = 433~cm^{-1}$) are in a good
agreement with the experimental data~\cite{Herzberg} ($R_e = 2.16~\AA$,
$\omega_e = 469~cm^{-1}$).  For the dipole moment we have obtained $D_e =
2.93~D$.

%*************************************************************************
\paragraph{Restoration of four-component spinor for valence electron.}
\label{S3a}

 In order to evaluate matrix elements of the operators singular near
 nucleus of barium, we have performed GRECP/SCF and GRECP/RASSCF
 calculations of BaF where the pseudospinors corresponding to $5s_{1/2}$,
 $5p_{1/2}$ and $5p_{3/2}$ shells were ``frozen'' with the help of the
 level-shift technique (that is also known as Huzinaga-type ECP,
 see~\cite{Huzinaga} and references therein). It was necessary to do
 because polarization of these shells were taken into account by
 means of EO technique (see below).
 Spin-orbit interaction was neglected for the explicitly treated electrons
 because of its smallness (see~\cite{TME}).  Thus, only core molecular
 pseudoorbitals occupying mainly atomic $1s, 2s$ and $2p$ orbitals of
 fluorine and the valence pseudoorbital of unpaired electron (that is
 mainly $6s,6p$-hybridized orbital of barium) were varied. RASSCF
 calculations with 5284 configurations were performed for 11 electrons
 distributed in RAS~1=(2,0,0,0), RAS~2=(2,1,1,0) and RAS~3=(6,4,4,2)
 subspaces.

 The molecular relativistic spinor for the unpaired electron was
 constructed from the molecular pseudoorbital $ \widetilde{\varphi}^M_u$
\begin{equation}
 \widetilde{\varphi}^M_u =
          \sum_i C^s_i \widetilde{\varphi}^s_i +
          \sum_i C^p_i \widetilde{\varphi}^{p,m_l=0}_i +
          \cdots,
 \label{3_1}
\end{equation}
 so that the atomic $s$- and $p$-pseudoorbitals of barium in~(\ref{3_1})
 were replaced by the unsmoothed four-component DF spinors derived for the
 same atomic configurations which were used in generation of basis
 $s,p$-pseudoorbitals.  The MO~LCAO coefficients were preserved after the
 RECP calculations.  As the spin-orbit interaction for the unpaired
 electron is small, the ``spin-averaged'' valence atomic
 $p$-pseudoorbital was replaced by the linear combination of the
 corresponding spinors with $j = l \pm 1/2$ (see~\cite{TME,Dmitriev}
 for details).

%*************************************************************************
\paragraph{Effective operators for valence electrons}

It is well known that the accuracy of the hyperfine structure calculations
for heavy atoms is not high if core polarization effects are not taken into
account. In \cite{IL} it was suggested, that correlations which are not
included in the active space, can be treated with the help of the EO. The
latter is constructed by means of the atomic many body perturbation theory
(for the application of the perturbation theory to the calculations of the
$P,T$-violation in atoms see, for example, \cite{Martensson}).  The main
advantage of this method is that there is no need to extend the active
space to include core electrons.

In \cite{IL} it was supposed that EO is constructed in the active space
which includes only few interacting levels. On the contrary, in
\cite{DFK,DFKP} it is suggested to use single EO for the whole (infinite
dimensional) valence space. Thus, all
correlations between valence electrons are treated explicitly, while EO
accounts only for the core excitations. In this case, EO is energy
dependent, but this dependence is weak if the energy gap between the core
and the valence space is not too small.  This makes EO method much more
flexible and allows to use one EO for different quantum systems, provided
that they have the same core.  In particular, it is possible to form EO for
the atom (or ion) and then use it in a molecular calculation.

Generally speaking, EO for the hyperfine interaction (as well as for any
other one-electron operator) is no longer one-electron operator,
even in the lowest order of the perturbation theory.  On the other hand,
the one-electron part of EO includes two most important correlation
corrections and in many cases appears to be a very good approximation. The
first correction corresponds to the Random Phase Approximation (RPA), and
the second one corresponds to the substitution of the Dirac-Fock
orbitals by the Brueckner orbitals.

To illustrate how EO works for the atomic barium, let us look at the
hyperfine constant of the $^3P_1 (6s6p)$-level of $^{137}$Ba. The
two-electron multiconfigurational Dirac-Fock calculation gives $A$ =
804~MHz \cite{ORFT}, which should be compared to the experimental value
1151~MHz.  The two-electron configuration interaction calculation with RPA
and Brueckner corrections included gives $A$~=~1180~MHz.

In this work we calculated EOs for the magnetic hyperfine
interaction, for the EDM of the electron and for the anapole moment. Both
RPA equations and Brueckner equations were solved for a finite basis set
in the $V^{N-2}$ approximation (which means that SCF corresponds to
Ba$^{++}$), and matrix elements of the EOs were calculated. The basis set
included Dirac-Fock orbitals for $1s \ldots 6s,6p$ shells. In addition
$7-21s,7-21p,5-20d$ and $4-15f$ orbitals were formed in analogy to the
basis set N2 of \cite{DFK}.  Molecular orbitals were reexpanded in this
basis set to find matrix elements of EOs for the molecular wave function.

%*************************************************************************
\paragraph{Results.} \label{S4}

Expressions for the electronic matrix elements which correspond to the
parameters $A$, $A_{\rm d}$ and $W_i$ of the operator \eref{1} can be found
in \cite{KL}. All radial integrals and atomic four-component spinors
were calculated for the finite nucleus in a model of uniformly charged
ball.

Results for the parameters of the spin-rotational Hamiltonian are given in
table~\ref{P,T-odd}. There are two measurements of the hyperfine
constants for $^{137}$BaF \cite{Knight,Ryz}. First of them was made for
a matrix-isolated molecule and second was performed in a molecular beam.
Results of these measurements were used in the semiempirical calculations
\cite{K85,KL} of $P$- and $P,T$-odd parameters of the
spin-rotational Hamiltonian. These calculations were based on the
similarity between electronic matrix elements for the hyperfine structure
interaction and for the $P$- and $P,T$-odd interactions. All of these
operators mainly depend on the electron spin density in the vicinity of the
nucleus. As a result, in a one-electron approximation parameters $W_i$ are
proportional to $\sqrt{A A_{\rm d}}$ \cite{K85}. Electronic correlations
can break this proportionality.

In table~\ref{P,T-odd} we give results of the SCF and RASSCF
calculations for 11 electrons with the restoration procedure described
above. It is seen that in these calculations parameters $A$ and $A_{\rm d}$
are significantly smaller than in experiments \cite{Knight,Ryz}.
On the next stage we used EOs to account for the core polarization effects.
That led to the 50\% growth for the constant $A$, while constant $A_{\rm
d}$ increased by 130\%.  Our final numbers for the hyperfine constants are
very close to the experiment \cite{Knight} (the difference being less than
5\%) but differ more significantly from \cite{Ryz}.

Our SCF and RASSCF results for all three constants $W_i$ are much
smaller than results of the semiempirical calculations \cite{K85,KL}. When
core polarization effects are taken into account with the help of
corresponding EOs, our values for $W_d$ and $W_{\rm A}$ dramatically
increase (at present we do not have RPA for the constant $W_{\rm S}$).
There is a good agreement between our final value for $W_d$ and that from
the semiempirical calculation, but for the constant $W_{\rm A}$, our
result is noticeably smaller.

It can be explained by the fact that proportionality between $W_d$ and
$\sqrt{A A_{\rm d}}$ holds within 10\% accuracy, but for the
constant $W_{\rm A}$ deviation from proportionality reaches 30\%.
Almost half of this deviation is caused by the finite nuclear size
corrections to radial integrals.
Electron correlation corrections for both constants are about 15\%.

Two conclusions can be made from the results of this work. First, as it
was suggested in \cite{TME}, core polarization effects play very important
role in calculations of parameters of the spin-rotational Hamiltonian for
heavy diatomic radicals. Second, results of the {\it ab initio}
calculations with core polarization included, are close to
the results of the semiempirical calculations, correlation corrections
being about 15\%.  The fact that two very different methods give similar
results confirms that it is possible to make reliable calculations for such
molecules.
%*************************************************************************

\paragraph*{Acknowledgments.}

 This work was supported in part by RFBR grant
% A.T., N.M. and P.S.
 N 96--03--33036a and RFBR/DFG grant 96--03--00069G.

%##########################################################################

\newpage
\narrowtext

\begin{table}
\caption{Excitation energies for low-lying states of Ba averaged over
nonrelativistic configurations (finite difference SCF calculations).}

\begin{tabular}{llddd}
&                     &  DF        &\multicolumn{2}{c}{GRECP}\\
\cline{3-5}
\multicolumn{2}{c}{Transition}
                      & Transition &  Absolute  & Relative \\
&                     & energy (au)&  error (au)& error (\%) \\
\hline
$6s^2 \rightarrow $
& $6s^1 6p^1         $ &  0.04813   &  -0.00003  &  0.06    \\
& $6s^1 5d^1         $ &  0.03942   &   0.00010  &  0.24    \\
& $6s^1              $ &  0.15732   &  -0.00002  &  0.01    \\
& $6p^1              $ &  0.24473   &   0.00002  &  0.01    \\
& $5d^1              $ &  0.18742   &   0.00017  &  0.09    \\
\end{tabular}
\label{Ba_conf}
\end{table}

\mediumtext
\begin{table}
\caption{Parameters of the spin-rotational Hamiltonian for BaF.}

\begin{tabular}{lddddd}
   & $A$  & $A_{\rm d}$ &   $W_d$    &    $W_A$    &   $W_S$  \\
   & ($MHz$) & ($MHz$) & ($10^{25}~\frac{Hz}{e~cm}$) & ($KHz$) & ($Hz$) \\
\hline
Exper.-I/Semiemp.\tablenotemark[1]
          &   2326.    &   25.    &   $-$0.41  &     240.    &  $-$13.  \\
Exper.-II/Semiemp.\tablenotemark[2]
          &   2418.    &   17.    &   $-$0.35  &     210.    &  $-$11.  \\
SCF
          &   1457.    &   11.    &   $-$0.230 &     111.    &   $-$6.1 \\
RASSCF
          &   1466.    &   11.    &   $-$0.224 &     107.    &   $-$5.9 \\
SCF/EO
          &   2212.    &   26.    &   $-$0.375 &     181.    &          \\
RASSCF/EO
          &   2224.    &   24.    &   $-$0.364 &     175.    &          \\
\end{tabular}
\tablenotetext[1]{Hyperfine structure constants measured for
matrix-isolated molecule \cite{Knight} and semiempirical calculation of
constants $W_i$ based on this experiment \cite{KL}.}

\tablenotetext[2]{Hyperfine structure constants measured for free
molecule \cite{Ryz} and semiempirical calculation
based on this experiment \cite{KL}.}

\label{P,T-odd}
\end{table}
\narrowtext

\end{document}